# Learning and Motivational Impact of Game-Based Learning: Comparing Face-To-Face and Online Formats on Computer Science Education

Daniel López-Fernández, Aldo Gordillo, Jennifer Pérez, Edmundo Tovar

*Abstract*—

*Contribution:* **This article analyzes learning and motivational impact of teacher-authored educational video games on computer science education and compares its effectiveness in both face-to-face and online (remote) formats. This work presents comparative data and findings obtained from 217 students who played the game in a face-to-face format (control group) and 104 students who played the game in an online format (experimental group).**

*Background:* **Serious video games have been proven effective at computer science education, however it is still unknown whether the effectiveness of these games is the same regardless of their format, face-to-face or online. Moreover, the usage of games created through authoring tools has barely been explored.**

*Research questions:* **Are teacher-authored educational video games effective in terms of learning and motivation for computer science students? Does the effectiveness of teacher-authored educational video games depend on whether they are used in a face-to-face or online format?**

*Methodology:* **A quasi-experiment has been conducted by using three instruments (pre-test, post-test and questionnaire) with the purpose of comparing the effectiveness of Game-Based Learning in face-to-face and online formats. A total of 321 computer science students played a teacher-authored educational video game aimed to learn about software design.**

*Findings:* **The results reveal that teacher-authored educational video games are highly effective in terms of knowledge acquisition and motivation both in face-to-face and online formats. The results also show that some students' perceptions were more positive when a face-to-face format was used.**

## I. INTRODUCTION

THE use of educational methodologies that positively impact student motivation, and consequently their academic performance, should be further explored [1]. Innovative methodologies that go beyond the traditional master class must be incorporated in order to increase students' engagement. As engineering degrees in general, computer science degrees usually have high dropout rates [2]. Moreover, when an online remote format (hereinafter, referred just as online format) is used in education, the dropout rate is even higher [3]. As a consequence, computer science education, both in face-to-face and online formats, needs motivating and effective learning methods to mitigate these high dropout rates.

A methodology that meets these needs is Game-Based Learning (GBL), which has caught the attention of educators and researchers over the last years due to its potential to motivate students and promote their learning [4]-[14]. However, more outcome data and analytics about GBL in engineering and computer science education is required [7], specially to explore the best conditions for adopting GBL in face-to-face and online formats.

Furthermore, teachers have difficulties to find video games aligned with the learning goals they are pursing [15], which undermine the GBL adoption. To avoid programming a video game aligned with the learning goals from scratch, it is critical to promote the usage of teacher-authored video games, which are defined as "educational video games created by teachers with authoring tools that allow them to create their own games without the need of writing code or strong computer skills".

Previous research works [13][14] carried out by the authors of this article have shown that this kind of games facilitates the implementation of GBL in face-to-face settings, as well as in online education, which has become even more relevant due to the Covid-19 pandemic [16].The first work [13] examined the usage of two teacher-authored educational video games on face-to-face computer science education, showing that these games were effective as traditional teaching in terms of knowledge acquisition, and that they successfully increased

This paragraph of the first footnote will contain the date on which you submitted your paper for review.

This work was supported in part by the *Universidad Politécnica de Madrid* through the educational innovation projects "Use of escape rooms and educational video games in higher education" (IE1920.6103) and "Creation, use and analysis of educational video games of various genres. (IE22.6102), and by the European Union and the Comunidad de Madrid through the project "eMadrid" (S2018/TSC-4307).

Daniel López-Fernández, Aldo Gordillo, and Jennifer Pérez are with the Computer Science Department, *Universidad Politécnica de Madrid*, Madrid, Spain (emails: daniel.lopez@upm.es, a.gordillo@upm.es, jenifer.perez@upm.es). Edmundo Tovar is with the Computer Languages and Systems and Software Engineering Department, *Universidad Politécnica de Madrid*, Madrid, Spain (emails: edmundo.tovar@upm.es).

student motivation. The second work [14] compared the effectiveness for online software engineering education of video-based learning and game-based learning using teacher-authored educational video games. The results showed that GBL using teacher-authored educational video games was more effective than video-based learning in terms of knowledge acquisition and motivation. Both works [13] [14] evidence the effectiveness of this kind of games in terms of motivation and knowledge acquisition in computer science education.

Despite these benefits, there is a lack of research about teacher-authored video games [17], which may be due to the lack of widespread use of effective and usable authoring tools [18]. These studies about such games [13] [14] [18] prove that it is still necessary to continue exploring the usage of these games to study in different settings whether they are effective in terms of learning and motivation.

Moreover, it is very common for teachers to combine face-to-face and online instruction modalities. In fact, most face-to-face courses have a virtual learning environment where teachers provide lecturers with digital resources such as slideshows and quizzes. In these courses, teachers can choose the format, face-to-face or online, in which to employ an educational video game. Therefore, it is necessary to find out whether the use of these games is more effective in face-to-face or online environments so that teachers could select the most appropriate format to use them.

To address these needs of studying and comparing GBL in face-to-face and online formats and its adoption with teacher-authored educational video games, this work poses the following two research questions:
- RQ1: Are teacher-authored educational video games effective in terms of learning and motivation for computer science students?
- RQ2: Does the effectiveness of teacher-authored educational video games depends on whether they are used in a face-to-face or an online format?

After analysing the answers to these questions, this work provides two main contributions to improve the knowledge about GBL in computer science education and teacher-authored educational video games. The first contribution is more evidence about the usage of educational video games in face-to-face and online formats in computer science education. And the second and premier contribution is new evidence on whether the format in which educational video games are used, face-to-face or online, affects their effectiveness.

To properly report these contributions and the chain of evidence that supports them, this article describes the conduction of a quasi-experiment. That description will facilitate its replication and extension in other scenarios and future researches. The quasi-experiment is supported by using three instruments: pre-test, post-test and questionnaire. It synthesizes the knowledge extracted from the analysis of the effectiveness of teacher-authored educational video games in both face-to-face and online formats as well as the comparison of the learning and motivational impact of GBL using teacher-authored games in face-to-face and online formats in a software engineering course.

The structure of the article is as follows. Section II presents the research methodology. Sections III depicts the results of the study and section IV discusses them. Finally, section V summarizes the conclusions and introduce further research.

## II. RELATED WORK

This article is focused on computer science education and the application of GBL by using teacher-authored educational video games, since they allow to customize the GBL experiences for specific educational areas, contents and learning objectives.

This is a necessary and consistent step forward from two previous works [13] [14] carried out by the authors of this article. As stated before, these works prove the effectiveness of teacher-authored educational video games in terms of motivation and knowledge acquisition, being even more effective than other traditional teaching methods. However, they did not compare the effectiveness of the educational processes performed in both face-to-face and online (remote) formats, which is the final purpose of the present article.

This related work section is organized in two parts: (A) Works related to GBL in engineering education; (B) Works related to comparing effectiveness of educational video games in face-to-face and online formats.

### A. GBL in engineering education

Literature reviews [4]-[11] show a wide range of studies that report empirical evidence on the positive contribution of GBL to student learning and motivation. Among these reviews, it is worth remarking the one performed by Bodnar et al. [7], since it is focused on the application of GBL in engineering education, including computer science. They concluded that the use of this methodology should continue to be extended because it is beneficial from a learning and motivational point of view.

Indeed, there are many studies that exemplify the attainment of these benefits in computer science education. For example, Malliarakis et al. presented a Multiplayer Online Role-Playing Game, named CMX, that proved to be useful for teaching programming both in terms of performance enhancement and motivation [19]. Another example is ScrumVR, a virtual reality video game that proved to be useful for teaching the Scrum methodology [20]. However, these videogames were created from scratch by teachers (who were also software developers), while video games like the presented in this contribution (i.e., teacher-authored video games) are created by teachers using authoring tools without the need of writing scripts or performing other actions that could require strong computer skills.

To the best of our knowledge, only this work [13] examined the usage of teacher-authored educational video games on face-to-face computer science education, while only these works [14] [18] examined them in online computer science education. In all cases, the above-mentioned learning benefits are reported: effective knowledge acquisition and motivation enhancement. In any case further research is needed to provide evidence of these results in different settings. Moreover, none of these studies [13] [14] [18] make a comparison of teacher-authored educational video games according to their format of usage: face-to-face and online.

### B. Comparing effectiveness of educational video games in face-to-face and online formats

Hartman et al. [21] addresses how to be efficient in face-to-face and online formats in education, however these formats are independently analyzed. The study underlines the benefits of GBL in a face to face format, as well as the use of games in online formats to increase the variety of activities aiming to promote the engagement. However, [21] does not include a comparison between GBL when conducted in face-to-face and online formats and their effectiveness.

Some recent studies like [22] and [23] compared the effectiveness of face-to-face and online formats on engineering education and pointed to some strategies to move from one format to another, thanks to multiple learning resources. Delving into different types of resources, the effectiveness of educational games in face-to-face and online formats has barely been compared and the existing researches are far from the engineering field, as for example [24], [25] and [26], which are focused on educational business games and report discrepant findings. While [24] and [25] suggest that educational games are more effective in online modality, [26] emphasizes additional benefits of the face-to-face format compared to the online one.

Two additional examples of this comparison can be found in educational escape rooms [27], [28]. They are quite different from educational video games, but, as they are ultimately a type of educational game, they are considered as part of the GBL. In the first example, authors study the effectiveness of an educational escape room for learning chemistry conducted in face-to-face and online modalities [27]. In the second one, authors explore and compare the learning and motivational effectiveness of an educational escape room for learning computer programming conducted in face-to-face and online modalities [28]. The findings of [27] suggest that face-to-face format provides with extra benefits and enhance some aspects such as engagement and motivation, whereas [28] indicates that learning effectiveness of the activity conducted online is lower than face-to-face. Nevertheless, it has not yet been compared whether teacher-authored educational video games are equally effective in face-to-face and online formats. As a consequence, teachers who combine both teaching formats cannot still be confident about which is the most convenient format to employ these games.

## III. RESEARCH METHODOLOGY

This article presents a quasi-experiment to rigorously study and compare the effectiveness of GBL in face-to-face and online formats. The next subsections describe the context and sample of the research, as well as the employed procedure, methods, instruments, materials, and data analysis techniques.

### A. Context and sample

This research is contextualized in a software engineering fundamentals course, which is a second-year mandatory course for most of the bachelor's degrees offered at the Faculty of Computer Systems Engineering at the *Universidad Politécnica de Madrid* (UPM), Spain. This course accounts for 9 ECTS credits and covers software development processes and methodologies, requirements engineering, software modeling, design, implementation, and testing. This research was specifically performed in a lesson about basic principles of software design, which are abstraction, modularity, information hiding, coupling and cohesion.

From a timeline perspective, this research was carried out during 2019-20 and 2020-21 academic years. During 2019-20 academic year, Covid-19 pandemic broke out and many activities such as education had to be carried out remotely. This circumstance provided us the online format scenario. Fortunately, during 2020-21 academic year, the pandemic started to subside, which provided us the face-to-face scenario.

The sample is represented by 321 students distributed as follows. The experimental group was composed by 104 students who were enrolled in the above-mentioned course during the academic year 2019-20 and took the lesson about software design in online format (hereinafter, the experimental group will be referred as the online group). The control group was composed by 217 students enrolled in the mentioned course during the academic year 2020-21 and who followed the course in face-to-face format (hereinafter, the control group will be referred as the face-to-face group).

Regarding breakdown of the sample by gender and age, the whole sample was composed by 273 men (85%) and 48 women (16%) with an average age of 20.7 (SD= 2.9). The online group was composed by 90 men (87%) and 14 women (13%) with an average age of 20.8 (SD= 2.7), whereas the face-to-face group was composed by 183 men (84%) and 34 women (16%) with an average age of 20.6 (SD= 3.1). The reported information is summarized in Table I.

TABLE I
SAMPLE STATISTICS

|  | Face-To-Face group | Online group | Total |
| --- | --- | --- | --- |
| Students | 217 | 104 | 321 |
| Men/Women | 183/34 | 90/14 | 273/48 |
| Average age | 20.6 | 20.8 | 20.7 |

### B. Procedure

The procedure for conducting the quasi-experiment consists of four steps. First, instructions for the activity were provided and a pre-test was distributed among all participating students to gauge their prior knowledge of the topic covered. Second, the students played for about 50 minutes the teacher-authored educational video game described in section II-D. Then, the students completed a post-test to effectively measure the knowledge attained through the educational video game. Lastly, after completing the post-test, the students completed a questionnaire to collect their opinion about the GBL experience.

All the required materials to complete this procedure were available on the Moodle platform of the software engineering fundamentals course. Access restrictions were configured at the Moodle platform to ensure that these resources were used in the presented order (e.g., until the pre-test is completed, a student cannot start the video game).

The students of the online group could do this activity at their own pace, as long as it was completed within a certain week of the course, whereas the students of the face-to-face group completed this activity during a face-to-face lesson.

## C. Methods and instruments

Pre-test and the post-test included the same 10 multiple choice-questions about the topic "basic principles of software design". These questions assessed students' knowledge in a theoretical-practical way. The students were given the same amount of time to complete the pre-test and post-test: 10 minutes for completing each test. Feedback was not provided to students after completing the pre-test to prevent them from memorizing the answers. Pre-test and post-test results did not count towards students' final grades to avoid cheating and other undesired behaviours. Pre-test and post-test were scored from 0 to 10.

Additionally, students had to complete a questionnaire which included two initial demographic questions (gender and age), as well as nine questions about their GBL experience that students were asked to agree or disagree using a Likert scale from 1 (total disagree) to 5 (total agree), a yes/no question, and an open question that required textual comments. For clarity, both the items of the questionnaire and its results, are presented together in the results section. The content validity of this questionnaire was checked by an expert revision, and its consistency and reliability were checked using the following statistics: the α of Cronbach and the Kaiser-Meyer-Olkin (KMO) coefficient [29], [30]. The former one is used to determine the consistency and reliability of the questionnaire and it resulted in 0.84, whereas the latter one is used to determine the sampling adequacy for the questionnaire and it resulted in 0.91. These results are considered positive because the α of Cronbach is above 0.8 (good level) and KMO is above 0.9 (excellent level) [29], [30].

## D. Educational materials

The material used during this GBL experience by both groups was an educational video game. This game was authored by a course teacher through the SGAME authoring tool [18]. It allows teachers to easily create educational web games by integrating SCORM-compliant learning objects into existing games. In the video game here presented, the learning objects were created by the course teacher through the ViSH Editor authoring tool [31], which is an online tool included in SGAME. In a nutshell, by using these tools teachers can create their own learning objects from scratch and integrate them on existing video games to make them educational. Detailed information on these tools and the step-by-step procedure to create educational video games can be found in [18].

The game created for this learning experience is based on a popular 2D mobile game called Flappy Bird. In this game the player controls a bird who continuously moves to the right between pairs of green pipes (see Fig. 1). The player can click on the screen or press the spacebar key so that the bird briefly flaps upward, which must be done if the player does not want the bird to fall to the ground because of gravity. The player gets points for each pair of pipes successfully crossed. When the bird touches a pipe a learning object is popped up. These learning objects are interactive presentations that contain a first slide which includes a self-grading multiple choice question (see Fig. 2), and subsequent slides which provide theoretical-practical concepts about the knowledge assessed by the posed question (see Fig 3). The player could continue to play as long as he/she answers the question correctly, but if the player fails then he/she will lose and will have to start over again. Although the playful purpose of the game is to get as many points as possible by circumventing pipes, the educative purpose is to consume learning objects correctly when colliding with a pipe. Both purposes are properly intermingled because to keep playing and get points, it is necessary for the player to answer the questions correctly. The more time passes the greater the pressure on the player, because once he/she has a high number of points he/she cannot afford to fail the question and start all over again.

The educational video game examined in this work was designed attending to the following criteria. First, the game should not have complex mechanics and any student, even those who do not usually play video games, should be able to quickly understand it and play it easily. Second, the game should allow the student to get more and more interested as the game progresses. Third, the learning objects integrated into the game should be popped up at an appropriate frequency and be of an appropriate size, so that the game mixes the ludic and didactic aspects that underlie any educational game.

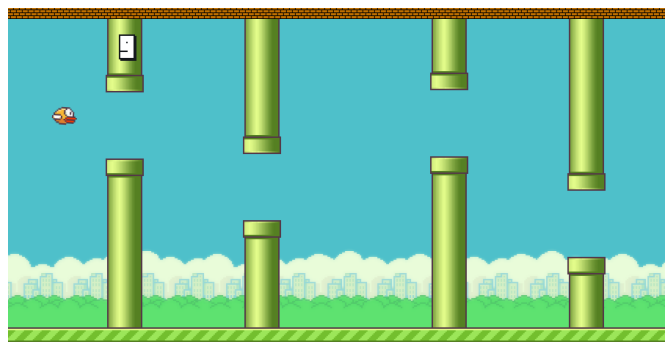

Fig. 1. Educational video game based on the Flappy Bird game.

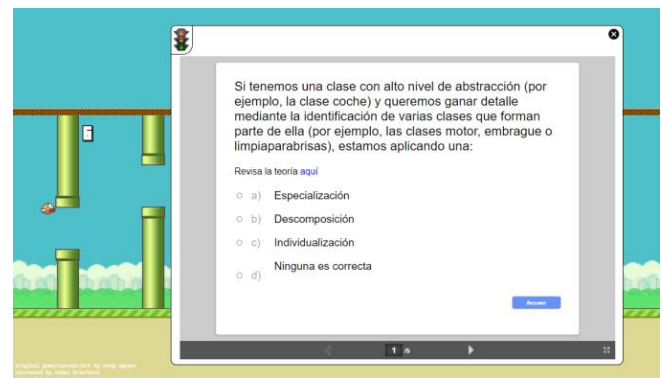

Fig. 2. Multiple choice question integrated into the educational video game.

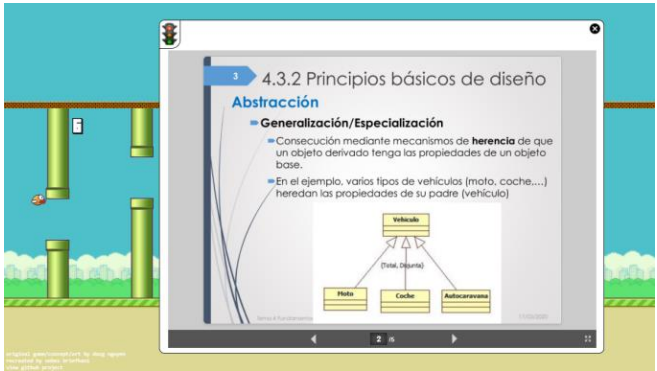

Fig. 3. Learning object integrated into the educational video game.

### E. Data analysis

First, normality of the data was checked by means of the Kolmogorov-Smirnov normality test. The results of this test indicated that the data were normally distributed. Therefore, parametric statistical methods were utilized. On the one hand, paired samples T-tests were performed to compare the differences between the scores achieved by the students of each group in the pre-test and the post-test. The magnitude of these differences is determined through the Cohen's d effect size [32]. On the other hand, independent samples T-tests were performed to compare the differences in the scores achieved in the tests and the learning gains between the online and face-to-face groups, as well as to compare the ratings obtained in the questionnaire by both groups. In these cases, Cohen's d effect size is also used to determine the magnitude of the differences. Lastly, it is noted that the thresholds indicated for Cohen's d effect size are as follows: the effect size is small from 0.2 to 0.49, medium from 0.5 to 0.79 and large from 0.8.

## IV. RESULTS

### A. Test results

Table II shows the pre-test and the post-test scores as well as the resulting student learning gains for the online and face-to-face group. For each score, the mean (M) and standard deviation (SD) is shown.

Regarding the learning gains (calculated as post-test score minus pre-test score) obtained by the participating students, statistically significant differences between pre-test and post-test scores were found in both groups using a T-test for paired samples. Both in face-to-face and online group, the difference has a medium to large effect size (Cohen's d $\geq$ 0.65). Therefore, it can be concluded that the GBL approach impacted positively and significantly on the acquisition of knowledge achieved by students in both the face-to-face and the online format.

Regarding the comparison between the two learning modalities, the scores of the two groups were compared by using T-test for independent samples. A slight difference was found in the pre-test scores, which were somewhat lower in the online group. However, that difference was not statistically significant, and its effect size was small (Cohen's d = 0.2). Another slight difference was found in the post-test scores, which in this case were somewhat higher in the face-to-face group. However, the difference was not statistically significant either and the effect size was not even small (Cohen's d = 0.15). Lastly, regarding the learning gains, these were slightly greater in the online format than in the face-to-face format, but this difference was not statistically significant, and its effect size was negligible (Cohen's d = 0.05). These results suggest that the educational video game had similar effectiveness in both formats in terms of knowledge acquisition.

TABLE II
RESULTS OF THE PRE-TESTS AND POST-TESTS

|  | Face-To-Face group (N=217) | Online group (N=104) | Independent samples T-test p-value | Cohen's d effect size |
|---|---|---|---|---|
| Pre-test [M (SD)] | 6.03 (1.77) | 5.69 (1.57) | 0.20 | 0.20 |
| Post-test [M (SD)] | 7.76 (1.66) | 7.51 (1.68) | 0.33 | 0.15 |
| Learning gains [M (SD)] | 1.73 (1.79) | 1.82 (2.03) | 0.74 | 0.05 |
| Paired samples T-test p-value | <0.001 | <0.001 | - | - |
| Cohen's d effect size | 0.71 | 0.79 | - | - |

### B. Questionnaire results

The items of the questionnaire used to examine the students´ perceptions about the GBL experience in both formats are presented in Table III. The results of this questionnaire are shown in Table IV.

The average rating for all items was 4.32 for the face-to-face group and 4.16 for the online group. In both groups the ratings given by students were quite high, although slightly higher in the face-to-face group. This can be appreciated in certain items where a statistically significant difference can be observed. In particular, a statistically significant difference with small to medium effect size (Cohen's d = 0.36) was found for overall opinion (item 1) and statistically significant differences with small effect size (Cohen's d $\geq$ 0.2) were found for self-reported learning effectiveness (item 2), fun (item 6) future use (item 8) and preference of GBL over the traditional learning approach (item 9). These results indicate that the face-to-face group have a more favorable perception of these key aspects for learning about the educational game than the students who did the activity in online format. However, the evaluation of other key aspects does not show statistically significant differences between the groups under study. In particular, no statistically significant differences were found in the items related to ease of use (item 3), required help (item 4), motivation (item 5) and integration with the learning platform (item 7). These results indicate that from the student's point of view, the educational game has a similar effectiveness in terms of motivation regardless of whether it is used in face-to-face or online format as well as it is fully usable in both formats.

When asked about their preference for the game-based learning method used over traditional learning methods based on videos or presentations, an overwhelming majority of students of both groups (97.2% in the face-to-face group and 94.2% in the online group) stated that they would not have preferred to learn about the targeted topic by using only presentations and videos. So, these results suggest that the surveyed students prefer educational video games to traditional materials.

TABLE III
ITEMS OF THE QUESTIONNAIRE

| Item | |
|---|---|
| 1 | My overall opinion about the educational video game is positive. |
| 2 | The educational video game helped me to learn. |
| 3 | The educational video game was easy to use. |
| 4 | I needed help to make use of the educational video game. |
| 5 | I found the educational video game engaging and motivating. |
| 6 | The educational video game made learning fun. |
| 7 | The educational video game was well integrated into the virtual platform from which I accessed it. |
| 8 | I would like to use educational resources similar to the educational video game used again. |
| 9 | I prefer to learn by playing educational games rather than using traditional materials (e.g. slide presentations). |
| 10 | Would you have preferred that we had taught this topic using only presentations and videos? |

TABLE IV
RESULTS OF THE QUESTIONNAIRE

| Item | | Face-To-Face group (N=217) | Online group (N=104) | Independent samples T-test p-value | Cohen's d effect size |
|---|---|---|---|---|---|
| 1 | M (SD) | 4.63 (0.58) | 4.37 (0.94) | < 0.01 | 0.36 |
| 2 | M (SD) | 4.59 (0.66) | 4.41 (0.84) | 0.01 | 0.25 |
| 3 | M (SD) | 4.76 (0.55) | 4.69 (0.73) | 0.28 | 0.11 |
| 4 | M (SD) | 1.76 (1.30) | 1.68 (1.35) | 0.57 | 0.06 |
| 5 | M (SD) | 4.37 (0.82) | 4.23 (0.98) | 0.14 | 0.16 |
| 6 | M (SD) | 4.38 (0.85) | 4.20 (0.99) | 0.05 | 0.20 |
| 7 | M (SD) | 4.66 (0.63) | 4.62 (0.77) | 0.63 | 0.06 |
| 8 | M (SD) | 4.68 (0.64) | 4.47 (0.85) | < 0.01 | 0.29 |
| 9 | M (SD) | 4.30 (0.96) | 3.98 (1.01) | < 0.01 | 0.39 |
| 10 | Yes (%) | 2.8% | 5.8% | - | - |

The comments gathered from the students were aligned with the presented quantitative results. Many comments pointed out the innovative character of the GBL experience or expressed gratitude to the teaching staff for their work (e.g., "*It is very appreciated these innovative activities to approach the subject. In my case this method makes me be more focused on the content than traditional classes*"), but most of them were referred to the fun and motivating aspect of the educational video game and its utility from a knowledge acquisition perspective (e.g., "*It was a very fun class, I had a great time but I also learned a lot*").

## V. DISCUSSION

*A. RQ1: Are teacher-authored educational video games effective in terms of learning and motivation for computer science students?*

The presented results allow to answer the RQ1 about the effectiveness of GBL using teacher-authored educational video games. The obtained learning gains point to the effectiveness of this learning method in terms of knowledge acquisition, whereas the questionnaire results reveal both quantitatively and qualitatively many beneficial aspects of GBL such as learning effectiveness, motivation, and fun. Moreover, the participating students clearly expressed that they prefer educational video games to traditional materials, at least for addressing the targeted topic in the described context. However, this would not necessarily be true if these video games were used in a different way, for example doing a weekly session of games would probably ending up boring them. Moreover, the preference of educational video games over traditional materials may not occur if the games were used by students from other educational levels (e.g., high school) or fields of knowledge (e.g., healthcare).

These results are consistent with the current body of scientific research about GBL [4]-[8] and research on specific video games designed in the computer education field [19][20], which underlines the positive impact in terms of learning and motivation of educational video games. In addition to providing further evidence on these benefits, this contribution reinforces the previous research about the usage of teacher-authored educational video games on computer science education in both face-to-face [13] and online [14] [18] scenarios and proves in two new educational settings related to computer science education that teacher-authored educational video games positively impact on student learning and motivation. Therefore, professionals who teach computer science, either in classroom or online format, can rely on this type of video games to complement their usual teaching.

*B. RQ2: Does the effectiveness of teacher-authored educational video games depends on whether they are used in a face-to-face or an online format?*

Regarding the effectiveness of GBL depending on the format of use, either face-to-face or online, the comparison of the learning gains obtained by the two groups indicates that both learning formats have a similar positive impact on the knowledge acquisition achieved by students. These are the premier results of comparing the use of teacher-authored educational video games in both face-to-face and online formats, since to the best of our knowledge, no study has previously compared students' knowledge acquisition achieved through this kind of games in both formats. When comparing these results with those obtained by using non-teacher-authored educational video games and other types of educational games like escape rooms, it can be evidenced that the results are not consistent with previous research [24], [25], [27], [28] (see section II.B).

Moreover, it deserves to be mentioned that the starting point (i.e., pre-test scores) of the online group was slightly lower. It should be noted that in the case of the online group, the activity was carried out at the beginning of the Covid-19 pandemic, and for several weeks there had been a halt in academic activity and some students may have personally suffered Covid-related problems. This could explain that the starting point of the online group was lower, and it is consistent with studies like [23], which points out the negative effects of the Covid-19 pandemic on student performance.

Furthermore, the questionnaire results indicate that some aspects of GBL are perceived more favorable in the face-to-face format than in the online one. These results indicate that the GBL experience conducted face-to-face outperformed the GBL experience conducted online in terms of the perceived learning effectiveness of the game, the fun achieved by playing it, and the overall opinion about the game, resulting in a greater likelihood of future use of this learning method and a greater preference for it over traditional methods. This may be because with the face-to-face format all students perform the activity at the same time and physically together, so the atmosphere is more festive and the fun is contagious. In addition, the feeling of competitiveness also increases and this, for many people, increases the fun. Therefore, despite the knowledge acquisition achieved with the educational video game through both formats is similar, the face-to-face format seems to enhance certain benefits of this modality.

These results are consistent with the reported on [26], which pointed out the additional benefits of the face-to-face format with respect to the online format as well as the importance of the facilitator in face-to-face settings. Moreover, our results are partially consistent with the reported on [27], which suggested that conducting an educational escape room in person rather than online provides extra benefits and enhance some aspects such as engagement and motivation. Lastly, our results partially differ from the reported on [28]. Unlike our findings, that study indicated that the level of fun, likelihood of future use of the activity and the preference for it over traditional methods did not vary from face-to-face format to online one, but that study agrees with our results on the fact that the overall opinion about the activity was more positive among those students who performed it face-to-face.

Furthermore, the obtained results indicate that the motivation achieved through both formats is very similar and the video game is highly engaging and motivating for students in online contexts too. Considering that online remote courses suffer from a large dropout rate of students, especially in science and engineering fields [3], this contribution is very valuable because it provides evidence on resources for improving student motivation in these settings.

Additionally, it was found that educational video games created using proper teacher-oriented authoring tools are easily usable not only in face-to-face format, but also online. Therefore, given the high importance that self-paced online learning settings are gaining, this finding is very interesting for many engineering educators who teach in online contexts who are eager to create their own educational video games for their courses.

Finally, given these findings, teachers who combine face-to-face and online formats can rely on using these video games in either format. Depending on their situation, teachers can choose to use a teacher-authored video game face-to-face or online with the confidence that both will be effective. For example, they can decide to take the face-to-face option if they need to infuse freshness into their face-to-face teaching or they can decide to take the online option if there are significant time constraints for conducting the GBL experience in person.

## VI. CONCLUSIONS

This article presents a quasi-experimental study to rigorously explore and compare the effectiveness of GBL in face-to-face and online formats when educational video games created by teachers through authoring tools are used in computer science education. In a nutshell, the results show that teacher-authored educational video games are highly and similarly effective in terms of knowledge acquisition and motivation in both formats. The comparative results indicate that the face-to-face format enhances some aspects of the learning experience such as the fun achieved by playing the game. However, the results also indicate that educational video games created with proper authoring-tools are fully usable and appropriate in online formats too and provide a similar level of learning and motivation to those of the face-to-face format. The obtained findings are summarized in Table V.

TABLE V. FINDINGS

| | |
|---|---|
| F1. | Teacher-authored educational video games are effective in terms of knowledge acquisition and motivation. |
| F2. | Teacher-authored educational video games have a similar positive impact on knowledge acquisition, regardless their format of use: face-to-face or on-line. |
| F3. | Some students' perceptions about GBL are similarly positive regardless of the format. For example, video games result highly engaging and motivating for students in both formats. |
| F4. | Some students' perceptions about GBL are more positive when the face-to-face format is employed. For example, video games result more fun when used in a face-to-face format thanks to the physical atmosphere of playing all students together. |
| F5 | There are no students' perceptions about GBL indicating that the online format outperformed the face-to-face one. |
| F6. | Educational video games created using proper teacher-oriented authoring tools are easily usable in both face-to-face and online formats. |

In addition, from these findings, it is possible to extract some recommendations for teachers. The findings F1-F3 evidence that the use of teacher-authored educational video games in both face-to-face and online formats is effective in terms of knowledge acquisition and motivation. As a result, it is possible to suggest that teacher-authored educational video games are recommended resources to facilitate the acquisition of knowledge as well as to promote the engagement in computer science education. Anyhow, students' perception is that a GBL experience is better in some aspects when conducted face-to-face (F4-5). As a consequence, it is recommended to incorporate mechanisms to enhance GBL online experiences such as leaderboards to promote competition or collaborative game mechanics to promote cooperation. Anyhow, since educational video games created by using teacher-oriented authoring tools are easy usable both in face-to-face and online format (F6), if there is no a video game that properly fits the learning goals of the subject, it is recommended to use a teacher-authored educational video game that meets the learnings goals pursued by the teacher.

Related with this, some recommendations to create these

games are provided:

1. Employ games with simple mechanics that can be easily mastered by the target students. If the mechanics are too complex, students unfamiliar with video games could experience difficulties to play it and hence could not address the desired educational objectives.

2. Design the game by properly balancing the ludic and the didactic aspects. For that purpose, the learning objects integrated into the game should be small to allow students to complete them in a short time (e.g., 1-2 minutes). Otherwise, the game will be interrupted for too long and students could get bored.

3. Trigger periodically the learning objects integrated into the game. They should be triggered with a frequency that is coherent with the size of the integrated learning objects (e.g., with a frequency of 30-60 seconds for learning objects that require 1-2 minutes to be completed). If the frequency is too high, the student will not have the feeling of playing a video game, and in the opposite case the learning effectiveness of the game will be low.

4. Design the learning objects integrated into the game by combining theoretical content and self-grading questions or more advanced self-assessment resources. Given that learning objects should be small, an interesting option is to include one single question together a small pill of information that contains the theoretical information necessary to answer such question.

In addition to recommendations for creating teacher-authored educational video games, some recommendations to conduct GBL experiences with these games are provided below:

1. Test the educational video game designed before executing a GBL experience in order to verify its correct operation and that the time scheduled for the experience is adequate.

2. Do not conduct too many GBL experiences during a course with the same students. In that case, the innovative effect of the games that brings freshness to the teaching process would be diluted and students might get bored. An appropriate number of GBL experiences for a semester course could be 1 or 2.

3. Design GBL experiences to last between 30 and 90 minutes. If the duration would be shorter, the didactic objectives of the game will not be successfully met and if they last longer, students may get tired of playing.

4. Include additional activities to enhance the learning effectiveness of a GBL experience. Examples of useful additional activities are an introductory lesson to lay the groundwork for the concepts to be covered during the game or a debriefing session to review the theoretical concepts covered throughout the game after playing it. It could be also interesting to include tests to be taken before and after the experience so that students become aware of the concepts they have acquired and those they have not.

On a different note, it should be commented that, in spite of the favorable results presented here, this study is not free of limitations. First, the research design is quasi-experimental as the compared groups took the activity in different years. Second, the study is focused on the evaluation of a single game. Nevertheless, it is evident that the use of teacher-authored educational video games provides important learning and motivational benefits in both face-to-face and online formats.

Lastly, future work is planned. First, to verify whether the benefits reported in the article go beyond computer science education and the stated findings are applicable in other contexts, further research exploring the usage of teacher-authored educational video games in more knowledge fields should be carried out. Second, to improve the GBL online experiences realized through video games created with the SGAME platform, mechanisms to promote cooperation (e.g., collaborative multiplayer games) and/or competition (e.g., leaderboards) should be implemented on the SGAME platform and validated by means of empirical experiences. Third, it would be interesting to compare the effectiveness of educational video games according to the game genre as it would help teachers to choose wisely the best type of game to be used with their students. Finally, another interesting future work would be conducting a longitudinal study to analyse with precision how GBL impacts on the dropout rates.

**DANIEL LÓPEZ-FERNÁNDEZ** received the Software Engineering degree and the Ph.D. degree in Software & Systems from the *Universidad Politécnica de Madrid* (UPM). He is currently Associate Professor with the Computer Science Department, UPM. His main research interests include the application of active learning methods, the development of innovative educational tools and the study of agile methodologies in software companies.

**ALDO GORDILLO** received the telecommunications engineering degree and the Ph.D. degree in telematics engineering from the *Universidad Politécnica de Madrid* (UPM). He is currently Associate Professor with the Computer Science Department, UPM. His research interests include the field of technology-enhanced learning, with a special focus on creation, evaluation, and dissemination of e-learning resources, computer science and software engineering education, game-based learning, gamification and e-learning systems.

**JENNIFER PÉREZ** is an associate professor at the E.T.S. Ingeniería de Sistemas Informáticos of the *Universidad Politécnica de Madrid* (UPM) since 2007. She received her PhD degree in Computer Science from the Polytechnic University of Valencia (UPV) in 2006 and is the recipient of the best thesis Award in Computer Science and Telecommunications of the UPV 2006-2007. Her research interests are focused on Systems of Systems, (SoS) Software Engineering, SPL, agile methodologies, game-based learning, gamification, open educational resources and educational innovation.

**EDMUNDO TOVAR** received the computer engineering degree and Ph.D. degree in informatics from the *Universidad Politécnica de Madrid* (UPM). He is currently a Professor with the Computer Languages and Systems and Software Engineering Department, UPM. Member of the IEEE Education Society Board of Governors (2005–2012), he is currently President elect (2019-2020) and President (2021-2022). He leads the REFERENT Research Group in technologies applied to Open Education. He is professional member of IEEE ETA KAPPA NU.